\documentclass[preprint,aps,nofootinbib]{revtex4}

\newcommand {\bc}{\begin{center}}
\newcommand {\ec}{\end{center}}
\newcommand {\bea}{\begin{eqnarray}}
\newcommand {\eea}{\end{eqnarray}}
\newcommand {\be}{\begin{equation}}
\newcommand {\ee}{\end{equation}}

\def\lsim{\mathrel{\rlap{\lower4pt\hbox{$\sim$}}
    \raise1pt\hbox{$<$}}}               
\def\gsim{\mathrel{\rlap{\lower4pt\hbox{$\sim$}}
    \raise1pt\hbox{$>$}}}   
\usepackage{graphicx}

\begin{document}


\title{From Nuclear to Unnuclear Physics}

\author{Thomas Sch{\"a}fer$^1$ and Gordon Baym$^2$}

\affiliation{
$^1$Department of Physics, North Carolina State University, Raleigh, NC 27695\\
$^2$Department of Physics, University of Illinois, 1110 W.~Green Street,  
  Urbana, IL 61801}

\begin{abstract}
We provide a brief commentary on recent work by Hammer and Son
on the scaling behavior of nuclear reactions involving the emission
of several loosely bound neutrons. In this work they discover  
a regime, termed {\em unnuclear physics}, in which these reactions
are governed by an approximate conformal symmetry of the nuclear force.
Remarkably, the scaling exponents that govern nuclear reactions
can be related to the energies of ultracold atomic drops confined
in harmonic potentials. We also comment on the importance and the
limitations of this approximate symmetry in the physics of neutron
stars.
\end{abstract}

\maketitle

\section{Introduction: Nuclear Physics}

Nuclear physics studies the stability, structure, and reactions of atomic 
nuclei, as well as the phase structure and properties of dense, strongly 
interacting matter.  While nuclear reactions are exceedingly complicated, 
remarkably Hammer and Son \cite{Hammer:2021zxb} have discovered a low energy
regime -- which they call {\em unnuclear physics} -- where certain nuclear 
reactions obey simple scaling rules, related to a new, emergent scaling symmetry 
of the nuclear force.   The term ``unnuclear'' is a reference to the work of 
Howard Georgi, who coined the term ``unparticle physics" to describe a possible 
scale invariant sector beyond the standard model of high energy physics 
\cite{Georgi:2007ek}.

  To appreciate the Hammer and Son result let us step back and recall some 
basics. Modern nuclear physics began with the discovery of the neutron in 1932, 
but ever since the 1970s we have known that the strong nuclear forces between 
neutrons and protons, collectively known as nucleons, are ultimately governed 
by Quantum Chromodynamics (QCD), the theory of quarks and gluons.  Quarks carry 
{\em color} charges, a generalization of electrical charge, and forces between 
``colored" particles are mediated by gluons. Neutrons and protons are in fact 
composite particles, and the nuclear force is the residual interaction between 
``color neutral" objects, similar to the van der Waals force between electrically 
neutral atoms and molecules. As a consequence, determining nuclear structure and 
reactions in detail requires numerical calculations based on complicated nuclear 
forces. Not only did Hammer and Son discover a simple tractable regime of nuclear 
reactions, even more remarkably, the dependence of the cross sections on energy 
that appear in this regime can be related to a completely different observable, 
the ground state energy of resonantly interacting ultracold atoms trapped in 
a harmonic potential. 

   Nucleon as well as nuclear reactions at low incident energies can be described 
in terms of two parameters, the ``scattering length," $a$, and the effective range, 
$r_e$, of the interaction.  The scattering length determines the low energy scattering 
cross section, $\sigma = 4\pi a^2$, and the effective range controls the leading 
energy dependence of $\sigma$.  In a quantum mechanical scattering reaction $a$ 
can be much larger than the physical size of the colliding particles or the range 
of the interaction potential between them. This happens, in particular, if the two 
colliding particles have an attractive interaction and are close to forming a bound 
state. Indeed, as the binding energy $B$ of the two particles approaches 0, the 
scattering length $a$ diverges. The effective range, on the other hand, is controlled 
by the physical range of the underlying interaction, and remains finite\footnote{
An essential feature of nuclear physics is the large separation of scales between 
the small binding energy of nuclei, typically a few MeV, and the large rest mass 
energy of the nucleon, about 940 MeV. An effective theory that exploits this 
separation of scales, constructed by Steven Weinberg in 1990 \cite{Weinberg:1990rz}, 
treats nucleons as point-like nonrelativistic spin 1/2 particles. In this theory, 
the nucleon interaction is controlled by long-range pion exchange, and zero-range
forces which are adjusted to reproduce the scattering parameters.}.

\section{Unnuclear Physics}  

 The limit in which the scattering length is much larger than all other 
length scales ($a\to\infty$), but the other length scales in the interactions 
are relatively small, is particularly interesting. In this case the physics 
is invariant under a rescaling of all distances \cite{Mehen:1999nd}.  
This phenomenon is well known in statistical mechanics, where in the vicinity 
of a second order phase transition the correlation length diverges, and 
fluctuations occur on all length scales. Correlation functions, for example 
the probability that a local density fluctuation in a fluid is correlated 
with a similar fluctuation separated by a spatial distance $x$, decay as a
fractional power of $x$. Striking physical phenomena, such as the critical 
opalescence of water near the liquid-gas critical end point, emerge as a 
consequence.   The milky appearance of the fluid arises from scattering 
of light by density fluctuations on a wide range of length scales. 

\begin{figure}[t]
\includegraphics[width=0.99\hsize]{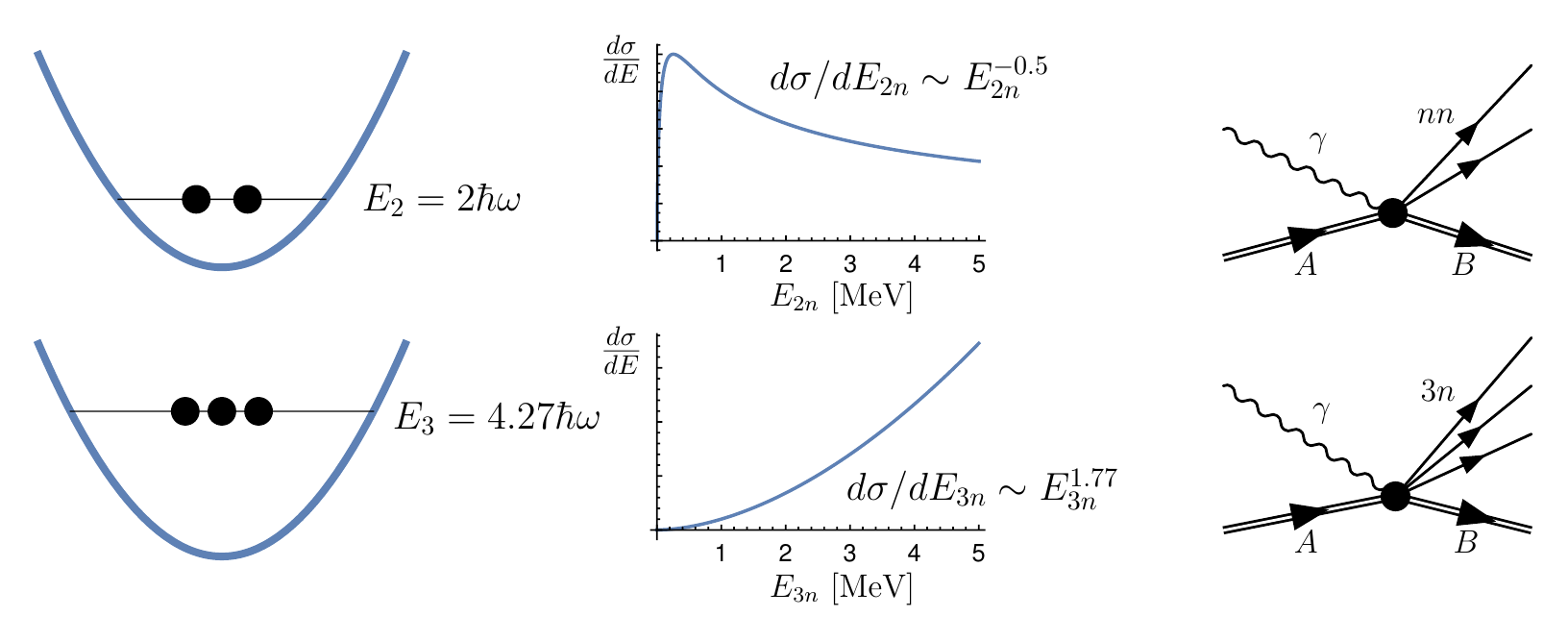} 
\caption{\label{fig:un} 
Relation between the ground state energy of resonantly interacting, harmonically
trapped atoms and the energy dependence of nuclear few-body reactions.  Hammer
and Son show that if the ground state energy of $N$ particles in a harmonic
trap is $E_N=\Delta\hbar\omega$ then the cross section for the reaction
$\gamma +A\to B+ Nn$, where an incident photon, $\gamma$, knocks out $N$ 
neutrons from a neutron-rich nucleus, $A$, scales as $E_{Nn}^{\Delta-5/2}$.
For two particles $\Delta=2$, and for three particles $\Delta=4.27$
\cite{Werner:2006zz}.}
\end{figure}

  In nuclear physics the neutron-neutron scattering length is very large, 
$|a_{nn}|\simeq 19$ fm (1 fm = $10^{-15}$ m), much bigger than the size of 
the neutron, of order 1 fm (or $10^{-15}$ m), so that a system of neutrons 
is close to the scale-invariant regime. The effective range is somewhat 
enhanced, $r_{nn}\simeq 2.3$ fm, but is significantly smaller than the scattering 
length. In a low energy reaction involving nucleons or nuclei, in which a 
momentum $q$ is transferred, with the length $\hbar/q$ lying well between 
the range of the nuclear forces and the scattering length, $r_{nn}\ll 
\hbar/q \ll a_{nn}$, the theory exhibits approximate scale invariance.  Such 
momentum transfers corresponds to a range of scattering energies, $\epsilon$, 
well between $\hbar^2/ma_{nn}^2$ and $\hbar^2/mr_{nn}^2$, or roughly
$0.1\,{\rm MeV} \ll \epsilon \ll 5\, {\rm MeV}$,  the regime of unnuclear 
physics. 

 Since finite nuclei are bound states, with discrete energy levels, it is 
not immediately obvious how to apply unnuclear physics to nuclear reactions.  
Hammer and Son observe that a natural arena for unnuclear physics is the 
disintegration of halo nuclei, neutron rich isotopes such as metastable 
$^{6}$He or $^{11}$Li that have several loosely bound neutrons. The cross 
section for disintegration of a halo nucleus by a low energy external probe, 
for example a photon, factorizes into a short distance piece characteristic 
of the particular nucleus, and a long distance part determined by unnuclear 
physics, the interaction between the neutrons in the final state. This implies
that the energy dependence of the cross section scales as a fractional
power of the total energy of the emitted neutrons, and that the exponent 
is determined by unnuclear physics. 

  This fractional power can be determined by methods similar to those 
developed in statistical mechanics \cite{Nishida:2007pj}. However, it is 
also possible to relate this exponent to a different physical problem.
There are no neutron bound states, but one can imagine a neutron drops confined 
in an external harmonic potential. Because of scale invariance the ground
state energy of the drop must be a multiple of $\hbar \omega$, where $\omega$ is 
the oscillator frequency. If the ground state energy of an $N$-neutron 
drop is $E=\Delta\hbar\omega$, where $\Delta$ is a positive number, the 
$N$-neutron unnuclear exponent is $\nu=\Delta-5/2$; see Fig.~\ref{fig:un}.

  The energy of neutron drops in a harmonic potential can be computed
by numerically solving the Schr\"odinger equation, but it cannot be 
measured experimentally. Experiments can be performed using a related 
system, ultracold fermionic atoms. These atoms can be tuned to the unnuclear 
regime using Feshbach resonances -- which modify the interaction strengths by 
means of external magnetic fields -- and be confined in harmonic potentials
using external laser fields. Experiments have succeeded in creating atoms 
drops with a controlled number of atoms \cite{Selim:2011}.

\section{From Unnuclear to nuclear matter}
  
  Another interesting subject, not directly addressed by Hammer and
Son, is the structure of unnuclear matter. In neutron matter 
at density $n=N/V$, where $V$ is the volume of the system, the typical 
interparticle separation $l$ is $\sim 1/n^{1/3}$. Thus the regime $r_{nn}\ll
l\ll a_{nn}$ is governed by unnuclear physics. Scale invariance implies
that up to a numerical factor the equation of state $P=P(n)$, where 
$P$ is the pressure, has to be equal to that of a free Fermi gas. 
This numerical factor, known as the Bertsch parameter $\xi$, can be
determined numerically, or using experiments with ultracold atomic
Fermi gases, which yield $\xi=0.37$ \cite{Ku:2011}. This 
result implies that the pressure of unnuclear matter is less than 
that of non-interacting neutron matter.  

  Unnuclear matter cannot be studied in the laboratory, but it does
exist in the inner crust of neutron stars. Neutron stars are compact 
objects with masses around 1-2 times the mass of the sun, and radii 
of about 12 km. The inner crust is a layer a few hundred meters below 
surface. The outer crust is a layer of nuclei embedded in a degenerate 
gas of electrons which provide the main source of pressure.  Once the 
neutron-rich nuclei in the crust can no longer accommodate further bound
neutrons, a neutron gas begins to permeate the crust.  At lower densities, 
this gas is in the regime of unnuclear matter. In the inner crust and 
in the outer layers of the core degenerate neutrons are the dominant 
source of pressure.  

  Observational constraints on neutron star masses and radii 
provide information on the equation of state of dense QCD matter.  
A central result is the discovery of neutron stars with 
masses in excess of two solar masses \cite{NANOGrav:2019jur}. In 
order to stabilize a star in this mass range matter at densities
beyond the unnuclear regime must be stiff, that means the pressure
as a function of density must increase more quickly than predicted
by the unnuclear equation of state. 

 This behavior is consistent with detailed calculations using 
nuclear forces \cite{Akmal:1998cf,Lonardoni:2019ypg}. In these
calculations the unnuclear behavior persists to densities of 
about 0.1 fm$^{-3}$, but at larger densities the pressure increases 
more quickly. This is caused by a combination of the neutron-neutron
force turning repulsive at larger momentum transfer, and repulsive 
three-body forces beginning to play a role. At even higher density,
greater than about 0.2 fm$^{-3}$, the description in terms of 
pointlike neutrons is no longer reliable, and quark degrees of freedom 
begin to play a role.

\section{Outlook}
  
 Hammer and Son compare their prediction to other theoretical 
calculations, but there are currently no experimental data in 
the unnuclear regime. This situation is expected to change in 
the near future, as part of the experimental program at radioactive
beam facilities such as RIKEN in Japan and FRIB in the U.S. 
Experiments with ultracold atoms will continue to explore few and
many-body systems of resonantly interacting atoms. An interesting 
problem is to understand non-equilibrium properties of unnuclear 
matter. Finally, observations at the gravitational wave observatory
LIGO, and the X-ray observatory NICER will pin down the mass-radius
relation of neutron stars, and constrain the equation of state beyond 
the unnuclear regime \cite{LIGO:2017vwq,Miller:2019cac,Riley:2019yda}.

Acknowledgments: Writing of this commentary was begun at the Aspen 
Center for Physics, which is supported by National Science Foundation 
grant PHY-1607611. T.~S.~also receives support from the US Department 
of Energy grant DE-FG02-03ER41260.


\end{document}